**Gene expression analysis reveals a strong signature of an interferon induced pathway in childhood lymphoblastic leukemia as well as in breast and ovarian cancer**.

Uri Einav[1], Yuval Tabach[1], Gad Getz[1], Assif Yitzhaky[1], Ugur Ozbek[3], Ninette Amariglio[2], Shai Izraeli[2], Gideon Rechavi[2,4] and Eytan Domany [1,4]

[1]Department of Physics of Complex Systems, Weizmann Institute of Science, Rehovot 76100 Israel, [2]Department of Pediatric Hematology-Oncology, Safra Children's Hospital, Sheba Medical Center and Sackler School of Medicine, Tel Aviv University, Tel Aviv, Israel, [3]Genetics Department, Institute for Experimental Medical Research (DETAE), Istanbul University, Turkey

[4]Correspondence: G. Rechavi, Department of  Pediatric Hematology-Oncology, Safra Children's Hospital, Sheba Medical
Center and Sackler School of Medicine,Tel Aviv University, Tel Aviv 52621,Israel.
email: gidi.rechavi@sheba.health.gov.il
E. Domany, Department of Physics of Complex Systems, Weizmann Institute of Science, Rehovot
76100 Israel. email: eytan.domany@weizmann.ac.il , phone: +972-(0)8-9343964

The abbreviations used are: ALL, acute ltmphoblastic leukemia; IIG, interferon inducible genes; CTWC, Coupled Two-Way Clustering; ATL, Adult T-cell leukaemia; HTLV, Human T-cell Lymphotropic Viruses; TNoM, Threshold Number of Misclassifications; EBV, Epstein-Barr virus; SPC, Superparamagnetic Clustering; FDR, false discovery rate; MMTV, mouse mammary tumor virus.

**ABSTRACT**

On the basis of epidemiological studies, infection was suggested to play a role in the etiology of human cancer. While for some cancers such a role was indeed demonstrated, there is no direct biological support for the role of viral pathogens in the pathogenesis of childhood leukemia. Using a novel bioinformatic tool, that alternates between clustering and standard statistical methods of analysis, we performed a "double blind" search of published gene expression data of subjects with different childhood ALL subtypes, looking for unanticipated partitions of patients, induced by unexpected groups of genes with correlated expression. We discovered a group of about thirty genes, related to the interferon response pathway, whose expression levels divide the ALL samples into two subgroups; high in 50, low in 285 patients. Leukemic subclasses prevalent in early childhood (the age most susceptible to infection) are over-represented in the high expression subgroup. Similar partitions, induced by the same genes, were found also in breast and ovarian cancer but not in lung cancer, prostate cancer and lymphoma. About 40% of breast cancer samples expressed the "interferon- related" signature. It is of interested that several studies demonstrated MMTV-like sequences in about 40% of breast cancer samples. Our discovery of an unanticipated strong signature of an interferon induced pathway provides molecular support for a role for either inflammation or viral infection in the pathogenesis of childhood leukemia as well as breast and ovarian cancer.

# INTRODUCTION

Recent years have witnessed accelerated improvement of gene expression measurement techniques, and a rapid growth of their usage, in particular for studies of malignancies. These technological advances were not accompanied by a similar rate of improvement in analysis methods (Ross et al., 2003; Yeoh et al., 2002). The present publication has two distinct aims. *First*, we demonstrate that when novel methods of analysis are applied to data that have been previously published and studied, it is possible to discover important molecular pathways that have completely eluded previous studies (Armstrong et al., 2002; Golub et al., 1999; Ross et al., 2003; Yeoh et al., 2002), that employed standard, commonly used methods for analysis of gene expression data. Our *second* goal is to present the discovery of a robust signature of a group of interferon inducible genes (IIG) associated with childhood leukemia and with other cancers, and to discuss its intriguing biological and clinical implications.

As has been discussed in several publications (Califano et al., 2000; Cheng & Church, 2000; Getz et al., 2000; Ihmels et al., 2002; Tanay et al., 2002), one of the main strengths of the modern gene expression technology also generates a considerable difficulty in interpreting the results. The strength is the holistic view achieved by measuring the expression levels of a very large number of genes in a single experiment. Typically, however, the expression signatures of an overwhelming majority of these genes are not related directly to the biological process (e.g. cancer) one wishes to study; in fact, most of the measured genes give rise to a very noisy background, from which one tries to extract the relatively weak signal of correlated activity of a small but relevant group of genes. A straightforward way to zero in on a relevant subset of genes is by means of a *supervised* filtering step – for example, identification of genes whose expression differentiates two or more groups of samples known to be genetically or clinically different. However, such a step can never lead to the discovery of unexpected partitions induced by genes whose role has not been previously anticipated. An alternative is provided by a family of methods (Califano et al., 2000; Cheng & Church, 2000; Getz et al., 2000; Ihmels et al., 2002; Tanay et al., 2002) that search for subgroups of genes and samples that satisfy certain conditions,, in an *unsupervised* manner. In particular, the Coupled Two-Way Clustering (CTWC, (Getz et al., 2000) clusters all genes as the first step, to identify correlated groups of genes; these gene-clusters are then used, one at a time, to probe and analyse the subjects. We use CTWC as our starting step, but deviate from this method in that the search is refined by a combination of more standard statistical tests (rather than continuing by unsupervised clustering), to zero in on an apparently interesting group of genes. Here we show that this mixture of supervised and unsupervised methodologies benefits from the advantages inherent to both methodologies and can lead to the discovery of biologically significant gene signatures.

The possible role of dysregulated immune or inflammatory response in the development of human cancer, and in particular its association with infectious agents, was suggested for many years. Several human lymphoid malignancies are associated with infectious agents: Burkitt's lymphoma with Epstein-Barr virus (Henle & Henle, 1966), ATL with HTLV-I (Poiesz et al., 1980), body-cavity lymphoma with human herpes 8 (Mele et al., 2003), B-cell non-Hodgkin Lymphoma with hepatitis C (Mele et al., 2003) and gastric MALT lymphoma with Helicobacter Pylori (Peek & Blaser, 2002).

It has long been suspected that common childhood infections contribute to the etiology of childhood leukemia, in particular ALL. The infectious etiology hypothesis has been proposed by two distinct but complementary theories. The Kinlen theory (Kinlen, 1995), based on transiently increased rates of leukemia in geographical clusters, suggests that population mobility and mixing result in infection occurring in susceptible, previously unexposed individuals. Several epidemiological studies supported the population mixing theory (Kinlen & Balkwill, 2001; Koushik et al., 2001). The alternative "delayed infection" hypothesis(Greaves, 1988; Greaves, 1997; Greaves & Alexander, 1993) focuses on the timing of common childhood infections and claims that some leukemia cases, mainly of the common B cell precursor subtype of ALL (cALL), are associated with a lack of exposure in infancy and a resultant failure of normal immune modulation. Dysregulated immune response upon delayed exposure to microbial infection is suggested to contribute to leukemogenesis. Studies in identical twins with leukemia (Ford et al., 1998; Wiemels et al., 1999), analysis of archived neonatal blood spots and screening of cord blood samples (Gale et al., 1997; Wiemels et al., 1999) indicate that cALL is frequently initiated by chromosomal translocations and non-disjunctions that occur prenatally, but requires a second "hit" to produce leukemia. The dysregulated response to infection is suggested to provide, probably indirectly, proliferative or apoptotic stress to the bone marrow, leading to the additional decisive "hit". The exposure is predicted to occur proximally to clinical disease, suggesting that a "smoking gun" can be identified when leukemia cell samples are studied. Despite intense research (MacKenzie et al., 2001; MacKenzie et al., 1999), no direct biological evidence, such as identification of microbial sequences, was found. Similarly, no epidemiologic data linking specific pathogens to ALL development were described. Several anecdotal reports described rare cases of ALL diagnosis preceded by a preleukemic phase known as pre-ALL in association with EBV or parvo B19 infection (Hasle et al., 1995; Tabori et al., 2001).

We describe here the identification of a gene expression signature in a subset of the patients suggestive of a deregulated immune response to some pathogen. We show that this signature occurs with highest frequency (one third) in the hyperdiploid ALL cases and as a smaller fraction in the other childhood leukemia subtypes. The finding of this gene expression profile in childhood leukemia, in particular in those cases that are overrepresented in the early childhood cALL peak, supports the role of an infectious agent, most probably a virus, in the pathogenesis of leukemia.

We looked for the same gene expression signature in a variety of data-sets of other human cancers. While in the majority of cancer samples no significant overexpression of the IIG was observed, it was detected in 40% of breast cancer and 20% of ovarian cancer samples. Indeed, some epidemiological studies have previously suggested a role for infection in the pathogenesis of ovarian (Ness et al., 2003) and breast (Ford et al., 2003) tumors. Additionally, molecular studies identified mouse mammary tumor virus-like sequences in about 40% of breast cancer samples (Ford et al., 2003; Wang et al., 1995). The "interferon signature" may reflect the activation of this pathway in the transformed cells themselves. Alternatively, it can reflect the response of the cancer cells to non-malignant cells of the immune system.

## RESULTS

**Analyzing the data of Yeoh et al** (Yeoh et al., 2002)**.**

Our aim was *class discovery:* to identify new partitions of the samples, into sub-groups with no previously known common label, on the basis of the expression profiles of a group of genes with correlated expression levels. To this end we used the CTWC method (see methods section). The expression levels of 3000 probe-sets that passed a variance filter were used in this analysis. We applied the algorithm on each of the ALL subtypes separately, in order to avoid 'inter-subtype' noise. ALL subtypes with large numbers of samples were the first to be analyzed. When we applied the algorithm on *TEL-AML1*, a group of 16 probe-sets representing 15 genes (Table 1) separated the *TEL-AML1* subtype very clearly into two sub-groups (Figure a): in 8 *TEL-AML1* samples these probe-sets had high expression levels whereas in the remaining 71 samples their expression level was relatively low. The distinct group of 8 samples shared no clinical label (such as same protocol of treatment or same prognosis). Strikingly, the majority (12 of 15) of the differentiating genes were interferon inducible genes (IIG). This constitutes *step (i)* of our analysis.

Next, in *step (ii),* we refined the list of these genes, using supervised analysis. We took the separation into the two groups of 8 versus 71 samples as "ground truth" and searched for genes that differentiate between these two groups. This search was performed on an extended set of 6500 genes. 184 probe-sets passed the TNoM as differentiating, with p-values below 0.05. To overcome the problem of multiple comparisons we applied the FDR method; 23 probe-sets, representing 19 genes, were identified as separating at an FDR level of 5%. The practical meaning of this statement is that out of these 23 probe-sets we expect about one to be a false positive, present due to random fluctuations.

The next *step, (iii),* of our iterative refinement process was again unsupervised; we used the expression levels of the 23 probe-sets found in step (ii), to characterize all samples, and clustered them using SPC. This way we identified a group of 50 samples, selected from all the ALL subtypes; these 50 have high expression levels of the 23 probe-sets (Figure 2). This group of samples consists mainly of *hyperdiploid>50* but contains almost all other subtypes as well (Table 2). The *hyperdiploid>50* subtype was significantly over-represented among the 50 samples with high expression; no other clinical label, specific to these samples, was found. Finally, to complete the refinement process, supervised analysis was performed again in *step (iv)*, using TNoM on 6500 probe-sets, revealing 28 genes that most significantly separate the new sub-group of 50 samples from the remaining 285 samples (see Table 1).

Each of the four steps yielded its own list of separating probesets (genes). Although not identical, these gene lists have significant overlaps, which can also be inferred from Table 1. Out of the total 30 of known genes (whose symbols are given in the Table), 17 are known to be induced by interferon; most of these have never been associated with leukemia.

**Analysis of other datasets** (Armstrong et al., 2002; Bhattacharjee et al., 2001; Golub et al., 1999; Ramaswamy et al., 2001; Rozovskaia et al., 2003; Shipp et al., 2002; Singh et al., 2002; Staunton et al., 2001; van 't Veer et al., 2002; Welsh et al., 2001; Welsh et al., 2001).

We now turned to search for other types of cancer in which a similar finding may hold; we tested whether we can find a sub-division of samples in other datasets on the basis of the expression levels of genes from the same pathway. However, in each of the following datasets we had to use a different subgroup of the separating genes, since some genes did not appear in these datasets and others had too many missing values. We ran the SPC algorithm for each dataset, using the appropriate subgroups of our gene list. Our aim was to find a distinct group of samples, in which these genes were overexpressed. In addition, we checked the sample labels in order to find common clinical indicators, shared by the members of the selected subgroup.

We applied the same method of analysis to the more recent leukemia data of the same group (Ross et al., 2003; Yeoh et al., 2002), where the Affymetrix HG-U133 microarrays, containing 45,000 probesets representing 33,000 genes, was used on a much smaller number of samples, 132 representative cases. Although the genes and their representation on these microarrays are different; we did find a subset of the IIG (14 genes) that appears on both chips and clearly identifies a small subgroup (18% of the samples) with high IIG expression levels. Among these the hyperdiploid>50 samples were very significantly over-represented.

We then turned to analyze the leukemia data (Table 3) of Golub et al.(Golub et al., 1999), Armstrong et al.(Armstrong et al., 2002) and Rozovskia et al.(Rozovskaia et al., 2003). In each of these datasets we also found clear subgroups (containing about 10% of the samples), with overexpressed levels of these genes. Again, no common label was shared by the subgroup members.

Next, we ran SPC on datasets of other types of cancer (Table 3): (lymphoma(Shipp et al., 2002), prostate(Singh et al., 2002; Welsh et al., 2001) various tumors (Ramaswamy et al., 2001; Staunton et al., 2001), ovary(Welsh et al., 2001), lung(Bhattacharjee et al., 2001) and breast(van 't Veer et al., 2002)). We found very small or negligible sub-groups of samples that co-expressed the unique sub-group of IIG in the lymphoma, prostate and lung cancers. Analysis of the data published by Ramaswamy et al.(Ramaswamy et al., 2001), which contains samples from various types of cancer, revealed a small sub-group, 7 out of 280, that also contains samples from other types of cancer, but mainly from leukemia, lymphoma and even from normal peripheral blood samples. In the lung cancer dataset of Bhattacharjee(Bhattacharjee et al., 2001), a clear separation of ~1.5% of the samples was detected. The most significant signal came from the breast cancer data of Van't Veer et al.(van 't Veer et al., 2002), where 40% of the samples overexpressed these genes (Figure 3), and the ovary cancer dataset of Welsh(Welsh et al., 2001), in which the interferon-related genes were overexpressed in about 20% of the samples.

**Confirmation of Differential Gene Expression by RQ-PCR**

For an independent verification of this bioinformatic analysis we have examined by RQ-PCR the expression of two of the interferon inducible genes IRIFT4 and IRF7 (table 1) in RNA derived from diagnostic bone marrow samples of 63 children with B cell precursor ALL. These patients were not part of the cohort included in the original microarray analysis of Yeoh and al. Despite the limitations imposed by the analysis of only two genes, using SPC, we have identified a cluster comprised from 10 patients with significantly higher expression of both genes. Interestingly, the average age of these patients was 4.45 years at the time of diagnosis, lower than 7.73, the average age in the low expression levels subgroup. The p-value for this age difference, assigned by the Student $t$ test, was $P = 0.011$. All patients but one in the small sub-group are in the age range of 2 to 6. There were no statistical significant differences in other clinical parameters (although this is a too small group to identify survival patterns). Thus, an analysis of gene expression by a different methodology (RQ-PCR) in an independent set of patients identified a similar cluster of interferon inducible genes, in a similar fraction (15.8%) of patients with B cell precursor ALL.

**DISCUSSION**

In this work we analyzed recently published gene expression data of different subtypes of childhood ALL by means of an unsupervised approach, using the SPC and CTWC clustering methods in order to search for a set (cluster) of genes, whose expression profile separates the samples into two (unanticipated) distinct groups. Such a gene cluster was found, and extended using the TNoM supervised method. The search for the characteristic gene set was performed in a totally unprejudiced "blind" way regarding either the separating gene set or the resulting partition of the samples. Surprisingly, a special set of genes was found to be highly expressed in a small minority (0-14%) of samples of the various leukemia subtypes, and in a relatively high percentage (37%) of the hyperdiploid (>50) ALL subgroup that constitutes a large part of the cases in the early childhood peak of leukemia. 17 out of the 30 known genes that appear in Table 1 are interferon inducible genes. These include signal transducer and activator of transcription 1 (STAT1) and interferon regulatory factor 7 (IRF7), both involved in signal transduction downstream to interferon receptors, as well as many interferon alpha induced proteins such as interferon induced protein 44, interferon induced transmembrane protein 3, interferon induced protein 35, 2'5'-oligoadenylate synthetase 1 and 2, myxovirus resistance 1 interferon-inducible protein 78 and adenosine deaminase RNA specific. Interferon gamma-induced proteins such as protein 30 and interferon gamma-induced transcription factor 3 were also found in the special gene cluster. Interestingly, several ubiquitin-conjugating enzymes such as E2L6 and E2A and proteasome system components such as activator subunit 2 (PA28 beta), some of them known to be induced by interferon, were also present in the cluster. Such proteins are involved in the generation of antigenic peptides that are presented to CD8+ T cells by MHC class I molecules. Taken together, many genes relevant to the immune response were found to be present in the special cluster of IIG that are highly expressed in the hyperdiploid leukemia variant. Of great interest is the presence of apolipoprotein B mRNA editing enzyme, catalytic polypeptide-like 3G (APOBEC3G) in the interferon related gene cluster. This enzyme was shown lately to confer antiretroviral defense against HIV and other retroviruses through lethal editing of nascent reverse transcripts)Mangeat et al., 2003; Zhang et al., 2003(. Hypermutation by editing mediated by this enzyme was shown to be an innate defense mechanism against retroviruses. One may speculate that the expression of this gene is an indication for retrovirus involvement in childhood leukemogenesis.

The existence of the IIG cluster in B-cell precursor childhood leukemia was confirmed in independent cohort using RQ-PCR. This RQ-PCR validation is preliminary and therefore the size of the examined gene set was limited. We plan to extend this analysis to a larger cohort with additional genes included in the IIG cluster.

The search for the IIG cluster in other datasets of several malignant diseases (Table 3) indicates that in other datasets of leukemia about 10% of the samples expressed the special gene set. In lymphomas, prostate, lung and datasets of a variety of tumors none or a very low percentage of the samples expressed the set. The exceptions are the dataset of 49 ovarian cancer samples and 96 breast cancer samples; A subset of ~20% of the ovarian cancers and ~40% of the breast cancers overexpressed the gene set. It is of interest that some epidemiologic studies suggested a role for infection in the pathogenesis of other cancers, in addition to leukemia, among them both ovarian (Ness et al., 2003) and breast (Ford et al., 2003) tumors. The lack of the IIG set in the majority of non-leukemic samples supports the significance of the finding in the leukemic samples.

In particular, the prominent appearance of hyperdiploid leukemic samples (that occur in early childhood, when viral infection is most likely to occur) and the significant lower age of the sub-group of patients from the independent cohort strengthen the hypotheses of Greaves and Kinlen.

The finding that about 40% of breast cancer samples displayed the 'infection associated' gene signature is of special significance. Retrovirus-like particles were demonstrated in a breast cancer cell line (Keydar et al., 1984) and MMTV-like gene sequences were detected by PCR in about 40% of breast cancer samples in several studies (Ford et al., 2003; Wang et al., 1995). Interestingly, the percentage of cases where retroviral gene sequences were identified is very similar to the percentage of cases where the interferon gene signature was identified (~ 40%). The experiment to be done is to look at the same tumor samples for both interferon-associated gene expression and for the MMTV-like gene sequences,

    The samples with high expression of the IIG constitute a minority of the malignant samples. In the leukemia hyperdiploid subgroup only one third of the samples were positive and in the breast cancer cases 40% were positive. Several explanations can be suggested for this finding. First, infection can represent only one type of causative factor or "second hit" event and other mechanisms may operate in the rest of the cases. Second, the role of infection may be indirect, via the dysregulated immune response. Under such a scenario the infectious agent can contribute to leukemogenesis in a transient "hit and run" fashion and its fingerprints may not be found at the time of diagnosis. In addition, it can be expected that in different populations the involvement of viruses can vary, due to environmental and genetic factors, as was recently suggested in the case of differential expression of MMTV-like sequences in breast cancer patients from Australian and Vietnamese origin (Ford et al., 2003).

    An immune response to viral infection is by no means the only reasonable explanation for the IIG signature discovered in these cancer samples. Since interferons are known to be produced by a variety of inflammatory cells (Colonna et al., 2002; Dalgleish & O'Byrne, 2002; Ernst, 1999) the induction of interferon responsive genes may reflect the degree of tumor inflammation. This may hold particularly for solid tumors, rather than leukemias. Tumor infiltrating lymphocytes are commonly found in breast and ovarian cancers (Georgiannos et al., 2003; Liyanage et al., 2002; Nzula et al., 2003; Reome et al., 2004). Thus the upregulation of

interferon inducible genes in a fraction of specific tumors may reflect the response of the cancer cells to interferon secreted by host immune cells. Since some of the genes presented in the IIG signature are associated with growth inhibitory properties (Chawla-Sarkar M, 2003; Sangfelt, 2001; Wall et al., 2003; Zhang et al., 2003) it is tempting to speculate that this signature may be associated with improved prognosis. Accordingly, hyperdiploid ALL is associated with the best response to chemotherapy and increased rate of apoptosis (Ito et al., 1999; Pui CH, 2004). Interestingly it has been recently demonstrated that the presence of intratumural T-lymphocytes correlated favorably with survival of patients with ovarian cancer (Zhang et al., 2003).

A minority of the genes in the IIG signature e.g STAT1 and IFIT4 may be induced by other cytokines or by retinoic acid or chemotherapy (Ihle & Kerr, 1995; Yu et al., 1997). However, most of the genes present in this signature are known to be induced prinipally by interferons. Also all the analyzed databases included only diagnostic samples prior to exposure to chemotherapy. Thus our finding, of a highly expressed "interferon cluster", combined with the epidemiological evidence, most likely implies an immune response, either to viral infection or to the tumor cells, leading to interferon secretion, activation of interferon receptors and STAT signaling, resulting in the activation of many interferon regulated genes. Nevertheless, another possibility, that the pathway was activated by a mutation in the cancer cells, independent of a response to the host environment , cannot be completely ruled out. Interestingly, three interferon receptor genes are located on chromosome 21, a chromosome that is always amplified in hyperdiploid leukemia (Heerema et al., 2000). The role of aberrant STAT signaling (mainly STAT3 but not the interferon induced STAT1) and constitutive STAT activation in leukemia is the subject of several recent publications (Benekli et al., 2003). It is unclear whether constitutive STAT activation itself is the cause or the result of a transforming process.

We have demonstrated that applying a novel blind unsupervised subgroup discovery approach to publicly available gene expression databases allows identification of previously unrecognized biologically meaningful molecular signatures. Specifically we identified a set of interferon- regulated genes characterizing mainly the hyperdiploid lymphoblastic leukemia, breast cancer and ovarian cancers (and, possibly, in other types of cancer that were not studied here). The various hypotheses raised by the finding of this novel gene signature in cancers can be tested experimentally by the research groups that published the original gene expression datasets. For example, it could be interesting to examine the interferon levels in stored serum from patients with childhood ALL or to correlate the presence of retroviral particles or the degree of infiltration of lymphocytes in breast cancer specimens with the interferon induced genes' signature, as well as searching for activating mutations or polymorphisms in interferon receptor genes in patients with hyperdiploid childhood ALL. Clearly, this finding generates several biologically testable hypotheses whose potential implications on diagnosis, therapy and prevention of childhood leukemia, breast cancer and other malignancies are evident.

## MATERIALS AND METHODS

**Patients and Specimens.**

**Microarray Data**. There are several publicly available gene expression datasets on leukemia (Armstrong et al., 2002; Golub et al., 1999; Rozovskaia et al., 2003; Yeoh et al., 2002) We analyzed the data of Yeoh et al.(Yeoh et al., 2002), that tested diagnostic bone marrow samples from 327 ALL patients using Affymetrix U95A microarrays containing 12,533 probe sets. The samples were collected at the time of discovery of the disease, prior to administering any therapy. Expression levels were measured for 335 samples of bone marrow and peripheral blood representing several different ALL subtypes (*T-ALL, E2A-PBX1, BCR-ABL, TEL-AML1, MLL, hyperdiploid >50* chromosomes*, hyperdiploid 47-50* and *hypodiploid*). We expanded the analysis to other publicly available datasets of leukemia (Armstrong et al., 2002; Golub et al., 1999; Rozovskaia et al., 2003) and other cancers including: lymphoma (Shipp et al., 2002), prostate (Singh et al., 2002; Welsh et al., 2001), ovary (Welsh et al., 2001), lung (Bhattacharjee et al., 2001) and breast (van 't Veer et al., 2002).

**Quantification by real time quantitative PCR (RQ-PCR):** The expression of IRF7 and IRFIT1 was quantified in RNA derived from diagnostic bone marrow samples given with an informed consent by 63 children with B cell precursor ALL. RNA isolation, cDNA synthesis and RQ-PCR were performed as described by us (U.O) before (Akyerli et al., 2005). For every sample the amount and the quality of RNA was normalized by dividing by the corresponding arrythmetic median of Beta 2 microglobin and c-Abl "housekeeping" genes. Primer sequences (5'-3') were: IRFIT1: forward CACATGGGCAGACTGGCAG, reverse GCGGAAGGGATTTGAAAGCT; IRF7 forward TCCCCACGCTATACCATCTACC, reverse CAGGGTTCCAGCTTCACCAG;

Beta–2-Microglobulin (B2M) forward TGCCGTGTGAACCATGTGAC, reverse ACCTCCATGATGCTGCTTACA; c-ABL forward CCCAACCTTTTCGTTGCACTGT, reverse CGGCTCTCGGAGGAGACGTAGA.

**Microarray data analysis**

**Preprocessing and filtering the data.** We worked with an expression matrix organized in 335 columns (samples) and 12,533 rows (genes). Each value in the matrix is the expression level of a certain gene in a certain patient. Rows (genes) in which more than 20% of the values were lower than some threshold (T=10) were removed. After this filtering 6,653 genes remained. In these rows the values that were lower than T were replaced by estimates based on the values of the 13 nearest neighbors' genes(Troyanskaya et al., 2001). Next, logarithm (base 2) of each entry was taken, and the genes were filtered on the basis of their variation across the samples. Two sets, of 3000 and of 6500 genes were chosen, on the basis of their standard deviations, for the Coupled-Two Way Clustering (CTWC) step and for the Threshold Number of Misclassifications (TNoM) test, respectively. Similar procedures were followed for each of the additional data-sets.

**Unsupervised analysis: Clustering.** In order to separate the ALL samples into unanticipated sub-groups, we searched for a cluster (e.g. correlated set) of genes with a distinct expression profile in one part of the samples, and another profile in the other part. Since hypothesis testing can not reveal unexpected partitions, unsupervised techniques, such as clustering, are more suited for such a task. *The CTWC method* (Getz et al., 2000) focuses on correlated groups of genes, one group at a time. Relevant subsets of genes and samples are identified by means of an iterative process, which uses at each iteration level stable gene and sample clusters that were generated at the previous step. The ability to focus on stable, statistically significant clusters that were generated by the underlying clustering operation is essential for the CTWC method. Since *Superparamagnetic Clustering (SPC)* (Blatt et al., 1996) provides a reliable stability index, it is the method of choice to use in the CTWC scheme. The SPC algorithm is based on the physical properties of inhomogeneous ferromagnets (Blatt et al., 1996; Blatt et al., 1997; Getz et al., 2000). The unsupervised CTWC step yields a list (cluster) of genes, whose expression levels separate the samples into two groups, which constitute the starting point of the next, supervised steps of the analysis.

**Supervised analysis**. We used supervised methods in order to expand and refine the list of genes that was obtained by the unsupervised CTWC step. Using hypothesis testing (TNoM)(Ben-Dor et al., 2000), we tested genes, one at a time, to see whether their expression differentiates the two groups of samples that were identified by CTWC. This step provides an extended set of genes, which is now used to identify, in an unsupervised manner, samples that belong to classes of relatively high expression. This procedure is reminiscent in spirit of the signature method (Ihmels et al., 2002), albeit the latter uses a known set of genes (or conditions) as it's seed and does not switch to supervised statistical tests to refine the genes it found.

For binary class comparisons we used a non-parametric statistical test, *TNoM* (Ben-Dor et al., 2000), which tests whether the expression value of a certain gene can predict the class of the sample. An informative gene is expected to have quite different values in the two classes, and thus we should be able to separate these by a threshold value. TNoM provides an appropriate score according to the quality of separation. For each score we calculate its P-value, as describe in Ben-Dor et al(Ben-Dor et al., 2000). In order to control contamination with false positive genes associated with multiple comparisons we used the method of Benjamini and Hochberg(Benjamini & Hochberg, 1995) that bounds the average false discovery rate (FDR); namely, the fraction of false positives among the list of differentiating genes.

**ACKNOWLEDGMENTS**

We thank Dr. Sema Sirma for her technical work and the patients and their families for their contribution. This project was supported by grants from the Israel Academy of Sciences and Humanities, the Ridgefield Foundation, the Minerva Foundation ,the Germany-Israel Science Foundation (GIF), the Istanbul University Reseach Fund and the Turkish Society of Hematology.  We are grateful to the Arison family for their donation to the Center for DNA Chips in the Pediatric Oncology Department, The Chaim Sheba Medical Center.

**Table 1. Genes that separate the ALL samples into two subgroups**

| Gene Probe ID | Title | Gene Symbol | TEL-AML1 CTWC step (i)[a] | TNoM 1 step(ii)[b] (P value) | TNoM 2 step(iv)[c] (P value) |
|---|---|---|---|---|---|
| 36927_at | chromosome 1 open reading frame 29 | C1orf29 | + | 0.000115 | 6.69E-35 |
| 925_at | interferon, gamma-inducible protein 30 | IFI30 | + | | 2.51E-32 |
| 915_at (32814_at) | interferon-induced protein with tetratricopeptide repeats 1 | IFIT1 | + | 6.06E-06 | 2.51E-32 |
| 37641_at | interferon-induced protein 44 | IFI44 | + | 6.06E-06 | 4.44E-31 |
| 37014_at | myxovirus (influenza virus) resistance 1 | MX1 | + | 6.06E-06 | 7.40E-30 |
| 38584_at | interferon-induced protein, tetratricopeptide repeats 4 | IFIT4 | + | 0.000115 | 1.17E-28 |
| 1107_s_at (38432_at) | interferon, alpha-inducible protein (clone IFI-15K) | G1P2 | + | 6.06E-09 | 3.41E-25 |
| 38389_at | 2',5'-oligoadenylate synthetase 1, 40/46kDa | OAS1 | + | 0.000115 | 4.43E-24 |
| 39263_at (39264_at) | 2'-5'-oligoadenylate synthetase 2, 69/71kDa | OAS2 | + | 0.000115 | 5.51E-23 |
| 38014_at | adenosine deaminase, RNA-specific | ADAR | | 6.06E-09 | 6.57E-22 |
| 38517_at | interferon-stimulated transcription factor 3, gamma | ISGF3G | | | 8.76E-19 |
| 1358_s_at | | | + | 6.06E-09 | 8.35E-16 |
| 38662_at | Homo sapiens, clone IMAGE:4074138, mRNA sequence | | | 6.06E-06 | 7.67E-15 |
| 37360_at | lymphocyte antigen 6 complex, locus E | LY6E | | 6.06E-06 | 3.94E-11 |
| 35718_at | SP110 nuclear body protein | SP110 | | | 2.35E-09 |
| 33339_g_at (32860_g_at) | signal transducer and activator of transcription 1, 91kDa | STAT1 | + | | 1.74E-08 |
| 464_s_at | interferon-induced protein 35 | IFI35 | | 0.000115 | 8.77E-07 |
| 914_g_at | v-ets erythroblastosis virus E26 oncogene like (avian) | ERG | | | 5.98E-06 |
| (36383_at) | | | | | |
| 40054_at | KIAA0082 protein | KIAA0082 | | | 5.98E-06 |
| 37352_at (3753_g_at) | **nuclear antigen Sp100** | SP100 | + | 6.06E-06 | 5.98E-06 |
| 34947_at (41472_at) | apolipoprotein B mRNA editing enzyme, catalytic polypeptide-like 3G | APOBEC3G | | | 3.97E-05 |
| 36845_at | nuclear matrix protein NXP-2 | NXP-2 | | | 3.97E-05 |
| 40505_at | ubiquitin-conjugating enzyme E2L 6 | UBE2L6 | | 0.000115 | 3.97E-05 |
| 41841_at | Homo sapiens clone 23718 mRNA sequence | | | | 0.000257 |
| 39061_at | bone marrow stromal cell antigen 2 | BST2 | | | 0.000257 |
| 32800_at | retinoid X receptor, alpha | RXRA | | | 0.000257 |
| 38805_at | TGFB-induced factor (TALE family homeobox) | TGIF | | | 0.000257 |
| 890_at | ubiquitin-conjugating enzyme E2A (RAD6 homolog) | UBE2A | | | 0.000257 |
| 40852_at | tudor repeat associator with PCTAIRE 2 | PCTAIRE2BP | + | | |
| 32775_r_at | phospholipid scramblase 1 | PLSCR1 | + | | |
| 36412_s_at | **interferon regulatory factor 7** | IRF7 | + | 2.36E-07 | |
| 41745_at | **interferon induced transmembrane protein 3 (1-8U)** | IFITM3 | | 6.06E-06 | |
| 676_g_at | 6-pyruvoyltetrahydropterin synthase | PTS | | 0.000115 | |
| 1184_at (41171_at) | proteasome (prosome, macropain) activator subunit 2 (PA28 beta) | PSME2 | | 6.06E-06 | |

[a] In the 'TEL-AML CTWC step (i)' column we mark by + the genes that were obtained by the initial CTWC (step (i) of our analysis); two out of the 16 probe-sets correspond to the same gene and hence only 15 genes are marked.

[b] The 'TNoM1 step (ii)' column gives P-values of the genes that separate 8 versus 71 *TEL-AML1* samples according to the TNoM test, at FDR=0.05. Only 19 genes are indicated (out of 23 probe-sets), again because of multiple representations. [c] The 'TNoM2 step (iv)' column indicates 28 probe sets that separate all the ALLs into two subgroups of 50 versus 285 samples (see text). The genes that are known to be part of the interferon-JAK/STAT pathway are in **bold face**. In cases when two probe sets represent the same gene symbol, the lower p-value was taken.

**Table 2**. **The number of samples from each subtype in the Yeoh et al.(Yeoh et al., 2002) dataset and in the new subgroup**

| Subtype name | Number of samples | Number in subgroup |
|---|---|---|
| Hyperdip>50 | 65 | 24 |
| TEL-AML1 | 79 | 10 |
| Pseudodip | 29 | 4 |
| Normal | 19 | 4 |
| Hyperdip 47-50 | 23 | 3 |
| T-ALL | 45 | 2 |
| BCR-ABL | 16 | 2 |
| MLL | 21 | 1 |
| Hypodip | 11 | 0 |
| E2A-PBX1 | 27 | 0 |
| Total: | 335 | 50 |
| | | |
| Novel subtype (as found by Yeoh et al.) | 14 | 6 |

**Table 3. Datasets that were examined by the CTWC algorithm using the interferon related set of genes**

| Datasets | Type of samples | Overexpressed samples |
|---|---|---|
| Golub et al.(Golub et al., 1999) | 60 **leukemic** samples of ALL and AML | 5 samples, 4 of them are MLL |
| Armstrong et al.(Armstrong et al., 2002) | 72 **leukemic** samples: 24 ALL, 20 AML and 28 MLL | 7 samples (4 ALL, 2 MLL, 1 AML) |
| Rozovskia et al.(Rozovskaia et al., 2003) | 60 **leukemic** samples of ALL, MLL and CD10- ALLs | 6 MLL and ALL |
| Shipp et al.(Shipp et al., 2002) | **Lymphoma** samples from 58 DLBCL patients and 19 FL patients | None |
| Welsh et al. (Prostate)(Welsh et al., 2001) | 55 **prostate** cancer samples | None |
| Singh et al.(Singh et al., 2002) | 102 **prostate** cancer samples | None |
| Staunton et al.(Staunton et al., 2001) | 60 samples from **various types of cancer** | Poor separation of 3 samples (taken from breast cancer, renal cancer and leukemia patients) |
| Ramaswamy et al.(Ramaswamy et al., 2001) | 280 samples from **various types of cancer** | 7 samples: 1 lymphoma samples, 2 leukemia AML samples, 1 breast cancer sample, 1 bladder cancer sample and 2 samples taken from normal peripheral blood |
| Welsh et al. (ovary)(Welsh et al., 2001) | 49 **ovary** cancer samples | ~10 samples |
| Bhattacharjee et al(Bhattacharjee et al., 2001) | 203 **Lung** cancer samples | 6 samples. Other ~40 samples expressed intermediate mRNA level. |

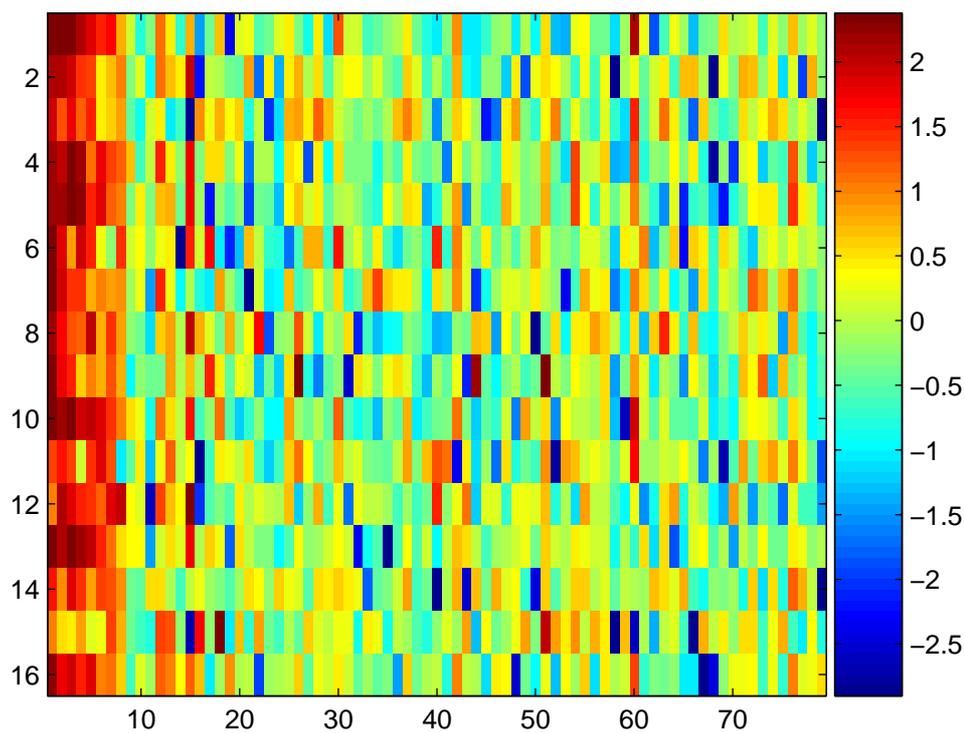

**Figure 1.** Expression values of the cluster of 16 genes, found by CTWC, in 79 *TEL-AML1* samples. The values are centered (mean expression of each gene = 0) and normalized (std = 1). These genes are overexpressed in a group of 8 (out of 79) TEL-AML1 samples.

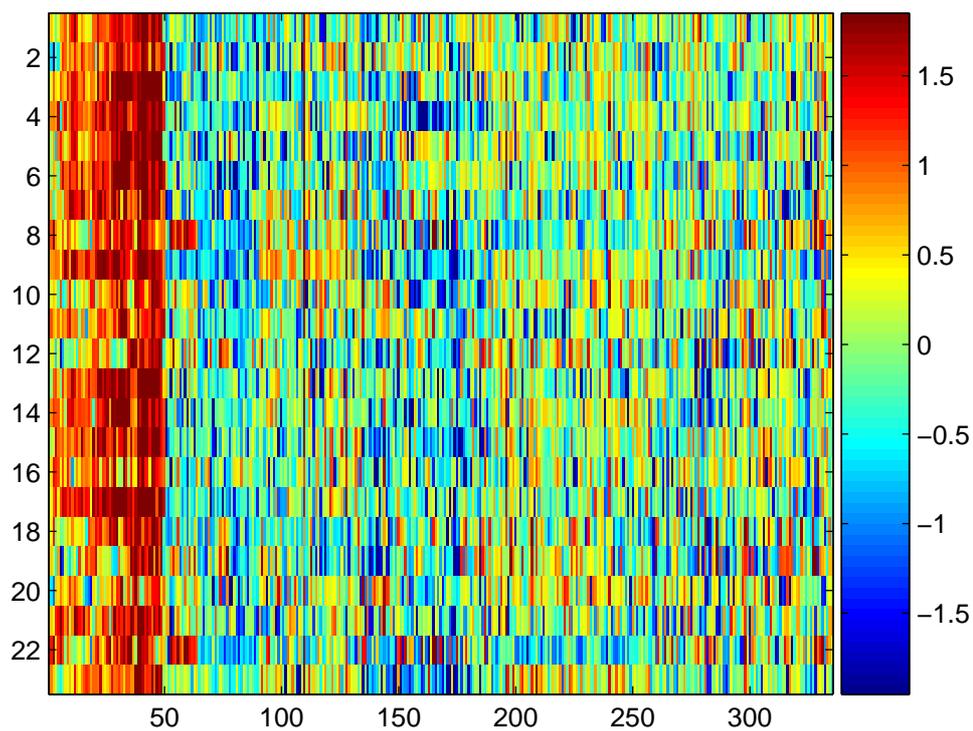

**Figure 2.** Expression values of the group of 23 genes in 335 ALL samples. The values are centered (mean expression of each gene = 0) and normalized (std = 1). These genes are correlated and overexpressed in a group of 50 ALL samples, that consist mainly of *hyperdiploid>50* and *TEL-AML1* subtypes.

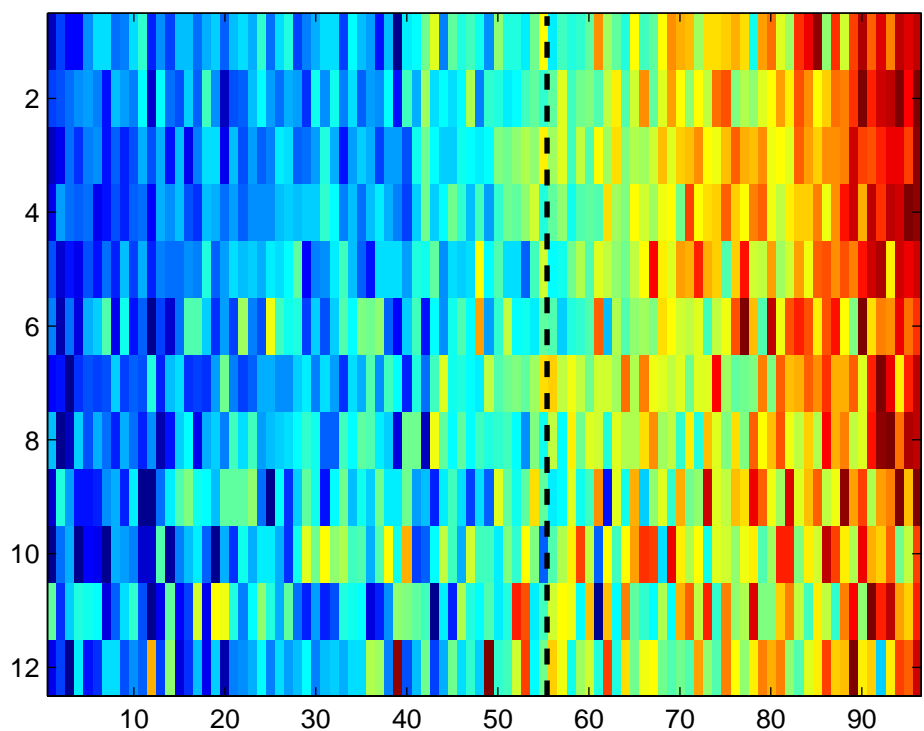

**Figure 3**. Expression values of 12 genes in 96 breast cancer samples (van 't Veer et al., 2002). The values are centered (mean expression of each gene = 0) and normalized (std = 1). Approximately 40% of the samples, to the right of the black dotted line, overexpress these genes. The differentiating genes are: IFI35, IFI30, STAT1, LY6E, OAS2, OAS1, IFIT1, UBE2L6, IFIT4, PLSCR1, IRF7, MX1.